\documentclass[12pt]{article}
\usepackage{graphicx}
\usepackage{epsfig, amsmath, amssymb}

\renewcommand{\bar}[1]{\overline{#1}}

\newcommand{\ket}[1]{\,\left|\,{#1}\right\rangle}

 \def\Pom{{ I\!\!P}}
 
 \def\gsim{\mathrel{\rlap{\lower4pt\hbox{\hskip1pt$\sim$}}
 \raise1pt\hbox{$>$}}}

 \newcommand\la{\langle}
 \newcommand\ra{\rangle}
 \newcommand\beq{\begin{equation}}
 
 \newcommand\eeq{\end{equation}}
 \newcommand\beqn{\begin{eqnarray}}
 \newcommand\eeqn{\end{eqnarray}}
\def\mb{\,\mbox{mb}}
\def\fm{\,\mbox{fm}}
\def\GeV{\,\mbox{GeV}}
\def\TeV{\,\mbox{TeV}}
\def\lsim{\mathrel{\rlap{\lower4pt\hbox{\hskip1pt$\sim$}}
    \raise1pt\hbox{$<$}}}         
\def\gsim{\mathrel{\rlap{\lower4pt\hbox{\hskip1pt$\sim$}}
    \raise1pt\hbox{$>$}}}         

\def\Im{\,\mbox{Im}\,}
\def\mb{\,\mbox{mb}}
\def\fm{\,\mbox{fm}}
\def\GeV{\,\mbox{GeV}}
\def\MeV{\,\mbox{MeV}}

\begin{document}
\begin{flushright}
USM-TH-181  \\
CPT-2005/P.021 \\
SLAC-PUB-11784\\
\end{flushright}
\bigskip\bigskip

\centerline{\large \bf Diffractive Higgs Production from Intrinsic
Heavy Flavors in the Proton}

\vspace{22pt}

\centerline{\bf {
 Stanley J. Brodsky\footnote{Electronic address:
sjbth@slac.stanford.edu}$^{a}$,
 Boris Kopeliovich\footnote{Electronic address:
bzk@mpi-hd.mpg.de}$^{b}$,
 Ivan Schmidt\footnote{Electronic address:
ivan.schmidt@usm.cl}$^{b}$,
 Jacques Soffer\footnote{Electronic address:
soffer@cpt.univ-mrs.fr}$^{c}$}}

{\centerline {$^{a}$Stanford Linear Accelerator Center,}
{\centerline {Stanford University, Stanford, California 94309, USA}}

{\centerline {$^{b}$Departamento de F\'\i sica, Universidad
T\'ecnica Federico Santa Mar\'\i a,}}

{\centerline {Casilla 110-V,
Valpara\'\i so, Chile}}

{\centerline {$^{c}$ Centre de Physique Th\'eorique, UMR 6207
\footnote{UMR 6207 is Unit\'e Mixte de Recherche
du CNRS et des Universit\'es Aix-Marseille I,\\
Aix-Marseille II et de
l'Universit\'e du Sud Toulon-Var - Laboratoire affili\'e \`a la FRUMAM.}
,}}
{\centerline {CNRS-Luminy Case 907, F-13288 Marseille Cedex 9, France}}

\vspace{10pt}
\begin{center}
{\large \bf Abstract}
\end{center}
 We propose a novel mechanism for exclusive diffractive Higgs
production $pp \to p H p  $ in which the Higgs boson carries a
significant fraction of the projectile proton momentum. This
mechanism will provide a clear experimental signal for Higgs production
due to the small background in this kinematic region. The key
assumption underlying our analysis is the presence of intrinsic heavy
flavor components of the proton bound state, whose existence at high
light-cone momentum fraction $x$ has growing experimental and
theoretical support. We also discuss the implications of this picture
for exclusive diffractive quarkonium and other channels.

\newpage

\section{Introduction}

A central goal of the Large Hadron Collider (LHC) being built at
CERN is the discovery of the Higgs boson, a key component of the
Standard Model, and whose discovery would constitute the first
observation of an elementary scalar field.  A number of theoretical
analyses suggest the existence of a light Higgs boson with a mass
$M_H \lesssim 130~\mbox{GeV}$.

In this paper we propose a novel mechanism for hadronic
Higgs production, in which the Higgs is produced with a significant
fraction of the projectile momentum.  The key assumption underlying
our analysis is the presence of intrinsic charm (IC) and intrinsic
bottom (IB) fluctuations in the proton bound
state~\cite{Brodsky:1981se, Brodsky:1980pb}, whose existence at high
$x$ as large as $x \simeq 0.4$ has a substantial and growing experimental and theoretical
support.   Clearly, this phenomenon can be
extended to the consideration of intrinsic top (IT).
A recent review of the theory and experimental constraints
on the charm quark distribution $c(x,Q^2)$ and its consequences for
open and hidden charm production has been given by
Pumplin~\cite{Pumplin:2005yf}. The presence of high $x$ intrinsic heavy quark components
in the proton's structure function will lead to Higgs production at high $x_F$
through subprocesses such as $g b \to H b$;  such reactions could
be particularly important in MSSM models
in which the Higgs has enhanced couplings to the $b$ quark~\cite{Nilles:1983ge}.

The virtual Fock state $|uud Q \bar Q>$ of a proton has a long
lifetime at high energies and can be materialized in a collision
by the exchange of gluons. The heavy quark and antiquark can then
coalesce to produce the Higgs boson at large $x_F \simeq x_c +
x_{\bar c}.$   This Higgs production process can be inclusive as
in $ p p \to H X$, semi-diffractive $p p \to H p X,$ where one of
the projectile protons remains intact, or exclusive diffractive $p
p \to p H p,$ where the Higgs can be reconstructed from the
missing mass distribution. In each case the Higgs distribution can
extend to momentum fractions $x_F$ as large as 0.8, reflecting the
combined momentum fractions of the heavy intrinsic quarks.

Perhaps the most novel production process for the Higgs is the
exclusive diffractive reaction, $pp \to p + H +
p$~\cite{DeRoeck:2002hk}, where the + sign stands for a large
rapidity gap (LRG) between the produced particles.  If both
protons are detected, the mass and momentum distribution of the
Higgs can be determined. The TOTEM detector~\cite{Deile:2004gt}
proposed for the LHC will have the capability to detect exclusive
diffractive channels. The detection of the Higgs via the exclusive
diffractive process $pp \to p + H + p$, has the advantage that it
does not depend on a specific decay mechanism for the Higgs. The
branching ratios for the decay modes of the Higgs can then be
individually determined by combining the measurement of $\sigma(pp
\to p + H + p)$ with the rate for a specific diffractive final
state $B_f~\sigma(pp \to p + H_{\to f} + p)$. This is in contrast
to the standard inclusive measurement, where one can only
determine the product of the cross section and branching ratios
$B_f~ \sigma( pp \to H_{\to f} X).$

The existing theoretical estimates for diffractive Higgs
production are based on the gluon-gluon fusion subprocess, where
two hard gluons couple to the Higgs $(gg \to
H)$~\cite{DeRoeck:2002hk}. A third gluon is also exchanged in
order that both projectiles remain color singlets. Perturbative
QCD then predicts $\sigma(pp \to p + H + p) \simeq 3~\mbox{fb}$
for the production of a Higgs boson of mass 120 GeV at LHC
energies, with a factor of 2 uncertainty \cite{DeRoeck:2002hk}.
Since the annihilating gluons each carry a small fraction of the
momentum of the proton, the Higgs is primarily produced in the
central rapidity region.

In this paper we will specifically consider the exclusive
diffractive production reaction $pp \to p + M +  p$, depicted in
Fig.~\ref{fi:1}, where $M$ stands for $J/\psi, \chi_c,
\Upsilon,\chi_b, Z ^0~\mbox{or}~ H$.  The final
state $M$ will be produced in the projectile proton's fragmentation
region with a significant fraction $x_F$ of the incident proton's
momentum, since the sum of the momenta of two heavy quarks contribute to the
momentum of $M$. This has an important advantage of providing a
distinctive signal with relatively small background. This
production process is analogous to the positron-antiproton
coalescence reaction by which anti-hydrogen was first
detected~\cite{Munger:1993kq,Baur:1995ck,Blanford:1997up}.

\begin{figure}[htb]
\begin{center}
\leavevmode {\epsfxsize=10.0cm \epsffile{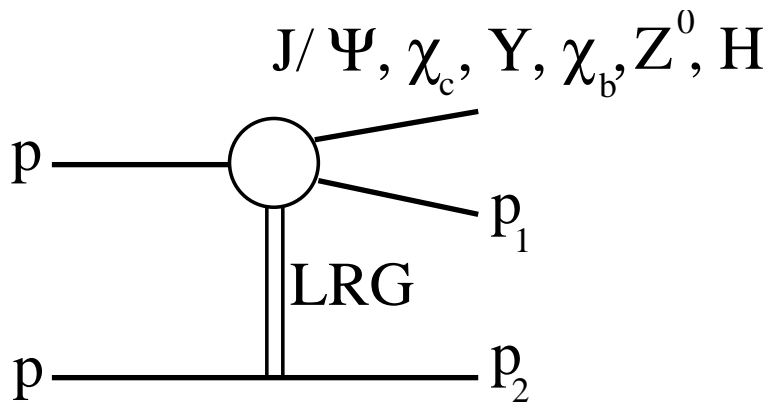}}
\end{center}
\caption[*]{\baselineskip 1pt The exclusive
diffractive production of
$J/\psi, \chi_c, \Upsilon,\chi_b, Z^{0} ~\mbox{or}~ H$, the Standard Model Higgs.}
\label{fi:1} \vspace*{-1.5ex}
\end{figure}

\section{Intrinsic Heavy Quarks}

The  proton eigenstate $\ket{p} = \sum_{n=3} \psi_n(x_i,\vec
k_{\perp_i},\lambda_i)  \ket{n; x_i,\vec k_{\perp_i},\lambda_i}$ of
the QCD Light-front Hamiltonian $H^{QCD}_{LF}$ can be expanded at
fixed light-front time $\tau=t+z/c$ as a superposition of quark and
gluon Fock eigenstates of the free Hamiltonian $H^0_{LF}: $  $\ket{p}
= \ket{uud}$, $\ket{uudg}, \cdots $, including, in particular, a
``hidden charm" Fock component $\ket{uudc\bar c}$. The fact that the
hadronic eigenstate has fluctuations with an arbitrary number of
constituents is a consequence of quantum mechanics and relativity.
The $\ket{uudc\bar c}$ Fock state arises in QCD not only from gluon
splitting which is included in DGLAP evolution, but also from
diagrams in which the heavy quark pair is multi-connected to the
valence constituents. The latter components are called ``intrinsic
charm" (IC) Fock components. The frame-independent light-front wave
functions $ \psi_n(x_i, \vec k_{\perp_i},\lambda_i) $ describe  the
constituents in the hadron with momenta $p^+_i = x_i P^+, \vec
p_{\perp i}= x_i \vec P_\perp +\vec k_{\perp i}$ and spin projection
$\lambda_i.$ Here $\sum^n_{i=1} x_i=1,$ and $\sum^n_{i=1} \vec
k_{\perp i}=\vec 0.$ The light-cone momentum fractions $x_i =
p^+_i/P^+ = (p^0_i+p^z_i)/(P^0+P^z)$   are boost invariant~\cite{Brodsky:1997de}.

It was originally suggested in
Refs.~\cite{Brodsky:1981se,Brodsky:1980pb} that there is a $\sim
1\%$ probability of IC Fock states in the nucleon; more recently,
the operator product expansion has been used to show that the
probability for Fock states in light hadron to have an extra heavy
quark pair of mass $M_Q$ decreases only as $\Lambda^2_{QCD}/M_Q^2$
in non-Abelian gauge theory \cite{Franz:2000ee}.  In contrast, in
the case of Abelian QED, the probability of an intrinsic heavy
lepton pair in a light-atom such as positronium is suppressed by
$\mu^4_{bohr}/M_\ell^4,$ where $\mu_{bohr}$ is the Bohr momentum.
The quadratic QED scaling corresponds to the dimension-$8$
Euler-Heisenberg effective Hamiltonian  $F^4/ M^4_\ell$ for light-by
light scattering mediated by heavy leptons.  Here $F_{\mu \nu}$ is
the electromagnetic field strength. In contrast, the corresponding
effective Hamiltonian in QCD $G^3/M^2_Q$ has dimension $6.$  This
difference in power behavior provides a remarkable discriminant
between non-Abelian and Abelian theory.

The maximal probability for an intrinsic heavy quark Fock state
occurs for minimal off-shellness;  i.e., at minimum invariant mass
squared ${\cal M}^2 = \sum^n_{i=3} (m^2_i +{\vec  k}^2_{\perp i})/
x_i  .$ Thus the dominant Fock state configuration is $x_i \propto
m_{\perp i}$ where $m^2_{i \perp} =m^2_i + {\vec k}^2_{\perp i};$
i.e.,  at equal rapidity.   Since all of the quarks tend to travel
coherently at  same rapidity in the $\ket{uudQ\bar Q}$ intrinsic
heavy quark Fock state, the heaviest constituents carry the largest
momentum fraction~\cite{Brodsky:1981se,Brodsky:1980pb}. Models for
the intrinsic heavy quark distributions $c_I(x,Q^2)$ and
$b_I(x,Q^2)$ predict a peak at $x \sim 0.4.$ Thus the intrinsic
heavy quarks are highly efficient carriers of the projectile
momentum.

Intrinsic charm also leads to new, competitive decay mechanisms for B decays
which are nominally CKM-suppressed~\cite{Brodsky:2001yt} and in
explaining the $J/\psi \to \rho \pi$ puzzle~\cite{Brodsky:1997fj}.
Furthermore, it has been found that intrinsic bottom could
even contribute significantly to exotic processes such as
neutrino-less $\mu^- - e^-$ conversion in nuclei~\cite{KSS}.

\section{Relevant Experimental Facts}

The most direct test of intrinsic charm is the measurement of the
charm quark distribution $c(x,Q^2)$ in deep inelastic lepton-proton
scattering $\ell p \to \ell^\prime c X.$ The only experiment which
has looked for a charm signal in the large $x_{bj}$ domain is the
European Muon Collaboration (EMC) experiment~\cite{Aubert:1982tt},
which used prompt muon decay in deep inelastic muon-proton
scattering to tag the produced charm quark. The EMC data show a
distinct excess of events in the charm quark distribution at $x_{bj}
> 0.3,$ at a rate at least an order of magnitude beyond predictions
based on gluon splitting and DGLAP evolution. Next-to-leading order
(NLO) analyses~\cite{Harris:1995jx} show that an intrinsic charm
component, with probability of order $1\%$, is needed to fit the EMC
data in the large $x_{bj}$ region. This value is consistent with an
estimate based on the operator product
expansion~\cite{Franz:2000ee}. Clearly it would be very valuable to
have additional direct measurements of the charm and bottom
structure functions at large $x$.

An immediate consequence of intrinsic charm is the production of
charmonium states at high $x_F = x_c + x_{\bar c}$ in high energy
hadronic collisions such as $pp \to J/\psi X $.  The $c$ and $\bar
c$ in the IC Fock state $\ket{uud c \bar c}$ can be materialized by
gluon exchange as a color-singlet pair which coalesces to a high
$x_F$ low $p_T$ quarkonium state.  The internal color structure of
the Fock state is important. The effective operator in the
non-Abelian theory predicts that the charm quark pair is dominantly
a color-octet $8_C.$ The color octet $(c \bar c)_{8_C}$ is then
converted to a high $x$ color-singlet $(c \bar c)_{1_C}$ state via
gluon exchange with the target; it then couples to the color-singlet
quarkonium state. Note that the  $J/\psi$ can be produced this way
only from the component of IC which is symmetric, relative to a
simultaneous permutation of spatial and spin variables.

Comprehensive measurements of the $p A \to J/\psi X$ and $\pi A
\to J/\psi X$ cross sections have been performed by fixed target
experiments, NA3 at CERN ~\cite{Badier:1983dg} and E886 at
FNAL~\cite{Leitch:1999ea}. According to the arguments in Refs.~
\cite{Brodsky:1989ex, Hoyer:1990us, Brodsky:1991dj}, the IC
contribution is strongly shadowed, thus accounting for the
observed nuclear dependence of the high $x_F$ component of the
$J/\psi$ hadroproduction. It is also important to consider effects
coming from energy conservation. Multiple interactions in a
nucleus can resolve higher Fock components of the  projectile
hadron compared to interaction with a free proton target.
Therefore, energy sharing between the projectile partons imposes
more severe restrictions on production of energetic particles
leading to nuclear suppression at large $x_F$
\cite{Kopeliovich:2005ym}."

The materialization of the intrinsic charm Fock state also leads to
the production of open-charm states such as $\Lambda(cud)$ and
$D^-(\bar c d)$ at large $x_F$.  This may occur either through the
coalescence of the valence and charm quarks which are co-moving with
the same rapidity, thus producing a leading particle effect or via
hadronization of the produced $c$ and $\bar c$. As shown in
Refs.~\cite{Barger:1981rx,Vogt:1995fs}, a model based on intrinsic charm
naturally accounts for the production of leading charm hadrons in $p
p \to D X$ and $p p \to \Lambda_c X$ as observed at the
ISR~\cite{ISR} and also at Fermilab~\cite{Aitala:2000rd,Aitala:2002uz}}. We also
note that it is also possible to construct Regge models which give
similar $x_F$ behavior as the IC approach.

The diffractive cross section $\sigma(p p \to \Lambda_c X + p)$, at
$\sqrt s=63~\mbox{GeV}$ was measured at the ISR to be is of order
$10$ to $60~ \mu b$~\cite{Giboni:1979rm}. This  result seems to be in
contradiction with findings of Fermilab experiments searching for
diffractive charm production~\cite{Wang:2000dq,e653}. The E690
experiment~\cite{Wang:2000dq} observed the diffractive channel
$\sigma(p p \to D^* p X) \sim 0.2\,\mu b$ at $\sqrt s =
40~\mbox{GeV}.$ Their results lead to the diffractive charm
production cross section $\sigma_{diff}(\bar cc)^{pp}= 0.6-0.7\,\mu
b$ which is about two order of magnitude smaller than the cross
section measured at ISR~\cite{Giboni:1979rm}. This result agrees
with the upper limit found for coherent diffractive production of
charm, $p Si\to \bar ccX + Si$, in the E653 experiment \cite{e653}
at Fermilab. However, forward charm production is most likely
strongly suppressed in a nuclear target as is the case for light
hadrons.  If one extrapolates to $pp$ collisions assuming an
$A^{1/3}$ dependence~\cite{kps}, the upper limit is
$\sigma_{diff}(\bar cc)^{pp}<7\,\mu b$. The ISR signals for forward
charm  production are thus not necessarily inconsistent with the
fixed target experiments considering the large differences in the
available center of mass energy, as well as the nuclear target
suppression.

It should be noted that diffractive charm production
via elastic scattering of the projectile plus gluon radiation also leads to the right order of
magnitude of the cross section. Indeed, the cross section for
diffractive gluon radiation (via the triple-pomeron term) in $pp$
collisions is known from data, $\sigma^{3\Pom}_{sd}\approx 4\mb .$
The production of a $\bar c c$ pair via a radiated gluon brings
extra factors from the coupling $\alpha_s\approx 0.2$ and from the
gluon propagator $(m_g/M_{\bar c c})^4$ where $m_g \approx 0.8
~\GeV$ is the effective gluon mass \cite{kst2}. Thus, one obtains an estimate for the
singly diffractive charm production cross section
$\sigma(p p \to c \bar c p X)
 \approx 4\,\mu b$, in good agreement with the magnitude of the data,
 at least in the central rapidity region.
However, the shape of the empirical $\Lambda_c$  distribution
at large $x_F$
is not readily accounted for this model.

One can produce the $\Lambda(bud)$ at high $x_F$ in inclusive $pp$ collisions,
through the materialization of the intrinsic bottom Fock state
$\ket{(uud)_{8_C} (b \bar b)_{8_C}}$.  The cross section for forward
open bottom production relative to open charm is reduced by the
relative IC/IB probability factor $m^2_c/m^2_b \sim 1/10$. Evidence for
the forward  production of the $\Lambda_b$ in $pp \to \Lambda_b X$ at
the ISR was reported in Ref.~\cite{basile}.

The existence of rare double-IC Fock state fluctuations in the proton, such as
$|uudc\bar{c}c\bar{c}>$ can lead to the production of two
$J/\psi$'s~\cite{Vogt:1995tf} or a double-charm baryon state at
large $x_F$ and small $p_T$.   Double-$J/\psi$ events at a high
combined $x_F \ge 0.5$ were in fact observed by
NA3~\cite{Badier:1982ae}. The observation of the doubly-charmed
baryon $\Xi^{+}_{cc}(3520)$ with mean $<x_F> \simeq 0.33$ has been reported  by the SELEX
collaboration at FNAL~\cite{Ocherashvili:2004hi};  the presence of two charm
quarks at large $x_F$ has, indeed, a natural IC interpretation.

\section {Intrinsic Heavy Quarks and Exclusive Diffractive Production}

We now investigate the implications of  IC and IB for  exclusive diffractive production
processes $ p p \to p M p$ at large $x_F$ where $M$ is a charmonium state, $Z^0$ boson, or the Higgs.  We first explain, using
Fig.~\ref{fi:2}, how the exclusive diffractive channels shown in
Fig.~\ref{fi:1} arise with the required color structure in the final
state.  As noted above, we shall assume that the projectile (upper)
proton has an approximate $1\%$ probability to fluctuate to an IC Fock component with the
color structure $|[uud]_{8_C}[\bar c c]_{8_C}>$.  This virtual state  has a long
coherence length in a
high energy collision $\propto {s/{\cal M}^2 M_p}$, where $\cal{M}$
is the total invariant final-state mass.  In a $pp$ collision, two soft gluons must be exchanged
in order to keep both protons intact and to create a
rapidity gap, mimicking pomeron exchange. The two gluons couple the  target nucleon to the
large color dipole moment of the projectile IC Fock state. For
example, as shown in Fig.~\ref{fi:2}, one of the exchanged gluon can be attached to
the $d$ valence quark spectator in $|[uud]_{8_C}[\bar c c]_{8_C}>$,
changing its color, and the other one can be attached to the $\bar c
$, also changing its color. The net effect of this color
rearrangement is the same as single-gluon exchange between the two
color-octet clusters. The $\bar c c$ and the $uud$ can thus emerge
as color singlets because of the gluonic exchange. The $[\bar c
c]_{1_C}$ can couple to the $J/\psi$, or to a $Z^0$ or to a $H$.
Meanwhile the color-singlet $uud$ gives rise to the scattered
proton, thus producing the two required rapidity gaps in the final
state.  Notice that the $x_F$ distribution of the produced particle
is approximately the same as the distribution of the $[\bar c c]$
inside the proton. As we shall discuss below, the sum of couplings of the gluon to all
of the quarks, as dictated by gauge invariance, brings in a form factor which vanishes at zero
momentum transfer, thus giving an
important suppression factor.

\subsection{The cross section}

\begin{figure}[tbh]
\includegraphics{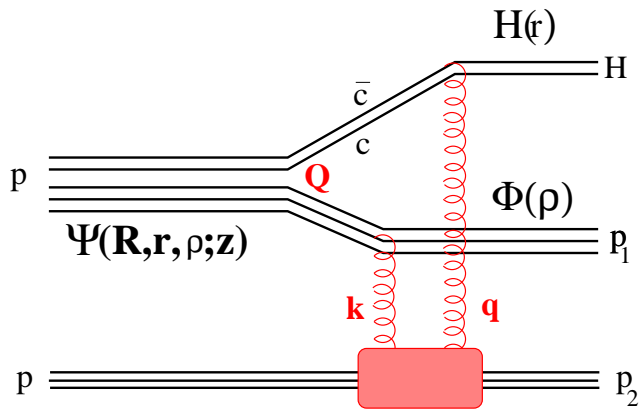}
\begin{center}
\vspace{6.5cm}
\parbox{13cm}
{\caption[Delta]
 { The two-gluon exchange diagram for the Higgs exclusive production }
 \label{fi:2}}
\end{center}
 \end{figure}

The cross section of exclusive diffractive production of the
Higgs, $pp\to Hp + p$, can be estimated in the light cone (LC) dipole approach~\cite{zkl}. The Born graph for this process is shown in
Fig.~\ref{fi:2}.
As discussed above, we shall assume the presence in the proton of an intrinsic charm (IC)
component, a $\bar cc$ pair, which is predominantly in a
color-octet state, and which has either a nonperturbative or
perturbative origin. In the former case this heavy component
can interact strongly with the $3q$ valence quark component.  Such nonperturbative
reinteractions of the intrinsic sea quarks in the proton wavefunction can lead to a $Q(x) \ne \bar Q(x)$ asymmetry as in the $\Lambda K$ model for the $s \bar s$ asymmetry~\cite{Burkardt:1991di,Brodsky:1996hc}.
As in charmonium, the mean $\bar cc$ separation
should be considerably larger than the transverse size $1/m_c$ of perturbative $\bar
cc$ fluctuations. For instance, if the binding potential has the
oscillator form, the mean distance is
 \beq
\la r_{\bar cc}^2\ra = \frac{2}{\omega\,m_c}\ ,
 \label{5}
 \eeq
 where $\omega\sim 300\MeV$ is the oscillation frequency.
Alternatively, the  IC component can be considered to derive
perturbatively from the minimal gluonic couplings of the heavy quark
pair to two valence quarks of the proton; this is likely the
dominant mechanism at the largest values of $x_c$~\cite{Brodsky:1991dj}.  In
this case the transverse separation of the $\bar cc$ is controlled
by the energy denominator, $\la r_{\bar cc}^2\ra = 1/m_c^2$, and is
much smaller than the estimate given by Eq.~(\ref{5}).

In accordance with the notation in Fig.~\ref{fi:2} the two protons in the CM frame
are detected with Feynman momentum fractions $x_1$ and $x_2$ and transverse
momenta $\vec p_1$ and $\vec p_2$ respectively. Correspondingly,
the produced Higgs carries  longitudinal momentum $ (x_2 - x_1)$ and transverse
momentum $\vec p_H=-(\vec p_1+\vec p_2)$. We assume the Higgs to
be heavy, over $100\GeV$, then $x_1$ and $x_2$ turn out to be
tightly correlated in this reaction. Indeed, the effective mass
squared of the $H-p_1$ pair reads,
 \beq
M^2=\frac{M_{H}^{2}+\vec p_H^{\,2}}{1-x_1}+\frac{m_{p}^2+\vec
p_1^{\,2}}{x_1}-\vec p_2^{\,2}\ .
 \label{7}
 \eeq
(We assume the equivalence of Feynman $x$ and the corresponding
fractions of the light-cone momenta, which is an  accurate
approximation at large $x$.)
Due to the form factors of the two protons, neither transverse
momenta, $\vec p_{1,2}$, can be much larger than a few hundred
$\MeV$ and therefore they, together with $\vec p_H$ and the proton
mass, can be safely neglected in Eq.~(\ref{7}). This could be incorrect
at very small values of $x_1\sim m_p^2/M_H^2$, but we will show
that the $x_1$ distribution sharply peaks at $\bar x_1\approx 0.25$.
Then, employing the standard relation $M^2/s=1 - x_2,$
 we arrive at the simple relation,
 \beq
(1-x_1)(1- x_2) = \frac{M_H^2}{s}\ .
 \label{250}
 \eeq

The diffractive cross section has the form,
 \beq
 \frac{d\sigma(pp\to ppH)}
{dx_2\,d^2p_1\,d^2p_2} = \frac{1}{(1- x_2)16\pi^2}\left|A(x_2,\vec
p_1,\vec p_2)\right|^2\ ,
 \label{10}
 \eeq
 where the diffractive amplitude in Born approximation reads,
 \beqn
A(x_2,\vec p_1,\vec p_2)&=&\frac{8}{3\sqrt{2}}\, \int
d^2Q\,\frac{d^2q}{q^2}\,
\frac{d^2k}{k^2}\,\alpha_s(q^2)\alpha_s(k^2)\, \delta(\vec q +\vec
p_2 +\vec k)\, \delta(\vec k -\vec p_1 -\vec Q) \nonumber\\
&\times& \int d^2\tau\,\left|\Phi_p(\tau)\right|^2
\left[e^{i(\vec k+\vec q)
\cdot\vec\tau/2} - e^{i(\vec q-
\vec k)\cdot\vec\tau/2}\right] \int
d^2R\,d^2r\,d^2\rho\,H^\dagger(\vec r)\ e^{i\vec q\cdot\vec r/2}
\nonumber\\ &\times&
\left(1-e^{-i\vec q\cdot\vec r}\right) \Phi_p^\dagger(\vec \rho)
e^{i\vec k\cdot\vec \rho/2} \left(1-e^{-i\vec
k\cdot\vec \rho}\right)\, \Psi_p(\vec R,\vec r,\vec \rho,z)\,
e^{i\vec Q\cdot\vec R}.
 \label{15}
 \eeqn
 Here $\Psi_p(\vec R,\vec r,\vec \rho,z)$ is the light-cone
wave function of the IC component of the projectile proton with transverse
separations $\vec R$ between the $\bar cc$ and $3q$ clusters, $\vec
r$ between the $c$ and $\bar c$, $\vec Q$ is the
relative transverse momentum of the $3q$ and $\bar cc$ clusters in
the projectile and $\vec\rho$ is the
transverse separation of the quark  and diquark which couple to the final-state proton $p_2.$
The density $|\Phi_p(\tau)|^2$ is the wave function of
the target proton which we also treat as a color dipole quark-diquark with transverse separation
$\tau$.  (The
extension to three quarks is straightforward \cite{zkl}). The
fraction of the projectile proton light-cone momentum carried by the $\bar cc$,
$z\approx 1-x_1$. This wave function is normalized as,
 \beq
 \int\limits_0^1 dz\int d^2R\,d^2r\,d^2\rho\,
\left|\Psi_p(\vec R,\vec r,\vec \rho,z)\right|^2
=P_{IC}\ ,
 \label{20}
 \eeq
where $P_{IC}$ is the weight of the IC component of the
proton, which is suppressed as $1/m_c^2$, and is assumed to be
$P_{IC}\sim 1\%$.   The amplitudes $H(\vec r)$ and
$\Phi_p(\vec \rho)$ denote the wave functions of the produced
Higgs and the outgoing proton, respectively, in accordance
with Fig.~2.

The phase factors in Eq.~(\ref{15}) correspond to different
attachments of the exchange gluons to quarks in Fig.~\ref{fi:2}.
Thus, attaching the gluon either to the $c$, or to the $\bar c$
quarks one gets factor $[\exp(i\vec q\cdot\vec r/2)-\exp(-i\vec
q\cdot\vec r/2)]$.  An analogous factor corresponding to the second
gluon coupling to the proton $p_1$ is also included in Eq.~(\ref{15}).
The transverse coordinates of the quark and diquark in the target proton
are $\tau/2$ and $-\tau/2$ (relative to its  center of gravity).
The phase factor in the square
brackets in Ref.~(\ref{15})  thus includes two terms corresponding to
attachment of the exchanged gluons to the same or different
valence quark or diquark in $p_2$.

In order to advance the calculations further, we will take the following
steps: First, we assume a factorized form of the proton wave
function,
 \beq
\Psi_p(\vec R,\vec r,\vec \rho,z) =
\Psi_{IC}(\vec R,z)\, \Psi_{\bar
cc}(\vec r)\,\Psi_{3q}(\vec \rho)\ .
 \label{1450}
 \eeq
 Here $\Psi_{\bar cc}$ and $\Psi_{3q}$ are the $\bar cc$ and
$3q$ wave functions normalized to unity, whereas
$\Psi_{IC}(\vec R,z)$ is the wave function describing the
relative motion of the $\bar cc$ and $3q$ clusters, where $z$ is the fraction of the longitudinal
momentum carried by the $\bar cc$. This wave function is
normalized as,
 \beq
\int d^2R\, \left|\Psi_{IC}(\vec R,z)\right|^2 =P_{IC}(z)\ ,
\label{1460}
 \eeq
 where $P_{IC}(z)$ is the $z$-distribution of $\bar cc$, related to the
$x_1$ distribution of the produced protons, since with a very high precision $z=
1-x_1=M_H^2/s(1-x_2)$ (unless $x_1$ is as small as $x_1\sim 2m_p/\sqrt{s}$).

We will perform the calculations in Eq.~(\ref{15}) only for
forward diffraction, i.e. $p_2=0$,
$\vec q=-\vec k$, and we assume for the Pomeron the typical
Gaussian $t$-dependence ($t=-p_2^2$),
 \beq
\frac{d\sigma}{d^2p_1\,d^2p_2}
 \propto e^{-B(s^\prime)p_2^2}\ ,
\label{1300}
 \eeq
so the $t$-integrated cross section then reads,
 \beq
\frac{d\sigma}{d^2p_1\,dx_2} =\left.\frac{\pi}{B(s^\prime)}\,
\frac{d\sigma}{d^2p_1\,d^2p_2\,dx_2}\right|_{p_2=0}\ .
\label{1350}
 \eeq
 Here the slope $B(s^\prime)\sim B_0 +
2\alpha_\Pom^\prime\,\ln(s^\prime/M^2_0)$, where
$B_0=4\GeV^{-2}$, $\alpha_\Pom^\prime=0.25\GeV^{-2}$,
$s^\prime/M_0^2=s/M_H^2$ and $M_0=1\GeV$.

The next step is to replace the two-gluon proton vertex,
represented by the integral over $\vec\tau$ in Eq.~(\ref{15}), by the
unintegrated gluon density, ${\cal F}(x,k^2)= \partial
G(x,k^2)/\partial({\rm ln}k^2)$, where $G(x,Q^2)=x\,g(x,Q^2)$.
This preserves the infra-red stability of the cross section, since
${\cal F}$ vanishes at $k^2\to 0$.  The phenomenological
gluon density fitted to data includes by default all higher order
corrections and supplies the cross section with an energy
dependence important for extrapolation to very high energies. One
can relate the unintegrated gluon distribution to the
phenomenological dipole cross section fitted to data for
$F_2(x,Q^2)$ from HERA, as was done in Ref.~\cite{gbw},
 \beq
 {\cal F}(x,k^2) =
\frac{3\,\sigma_0}{16\,\pi^2\,\alpha_s(k^2)}\ k^4\,R_0^2(x)\,
{\rm exp}\Bigl[-{1\over4}\,R_0^2(x)\,k^2\Bigr]\ .
 \label{1410}
 \eeq
 The problem is the extrapolation to the small virtualities $k^2$ typical
for the process under consideration.  The Bjorken variable is not a proper
variable for soft reactions; therefore we use the parametrization
from Ref.~\cite{kst2} adjusted to data for soft interactions.
Then $R_0(x)$ in Eq.~(\ref{1410}) should be replaced by $R_0(s^\prime)
= 0.88\,\fm \times(s^\prime/s_0)^{-\lambda/2}$ with
$\lambda=0.28,$ $s_0=1000~ \mbox{GeV}^2$ and $\sigma_0 \Rightarrow
\sigma_0(s^\prime)=\sigma^{\pi p}_{tot}(s^\prime)\, [1 +
3R^2_0(s^\prime)/8\la r^2_{ch}\ra_{\pi}]$, where $\sigma^{\pi
p}_{tot}(s^\prime)=23.6\mb\times(s^\prime/s_0)^{0.08}$ is the Pomeron
part of the $\pi p$ total cross section. The energy variable
$s^\prime$ is related to the rapidity gap between the two protons
in the final state, controlled by $x_2$,
 \beq
 s^\prime=
\frac{M_0^2}{1 - x_2}=s\,(1-x_1)\, \frac{M_0^2}{M_H^2}\ .
\label{1420}
 \eeq

Finally, combining all the above modifications and performing the
$p_1$-integration in Eq.~(\ref{15}), we arrive at,
 \beqn
\frac{d\sigma^{IC}(pp\to ppH)}{dx_2}&=&
\frac{32\pi\,P_{IC}(z)}{9\,B(s^\prime)(1- x_2)} \left|\int
\frac{d^2k}{k^4}\,\alpha_s(k^2)\, {\cal F}(x,k^2)\,
\right.\nonumber\\
&\times & \left. \int d^2r H^\dagger(\vec r)
e^{-i\vec k\cdot\vec r/2} \left(1-e^{i\vec k\cdot\vec r}\right)\,
\Psi_{\bar cc}(\vec r)\right. \nonumber\\
&\times&\left. \int d^2 \rho \,\Phi_p^\dagger(\vec \rho) e^{-i\vec
k\cdot\vec \rho/2} \left (1-e^{i\vec k\cdot\vec \rho}\right)\,
\Psi_{3q}(\vec \rho) \right|^2\ .
 \label{1400}
 \eeqn
 Here
 \beq
z=[x^H_F+\sqrt{(x_F^H)^2+4M_H^2/s}]/2\approx x_F^H\approx 1-x_1=
M_H^2/s(1-x_2)\ .
\label{1405}
 \eeq
This relation receives sizeable corrections only at very small Higgs
Feynman $x^H_F\sim 2M_H/\sqrt{s}$.
Notice that the expansion of the exponentials in
Eq.~(\ref{1400}) contains only odd powers of $\vec k\cdot\vec r$ and $\vec
k\cdot\vec \rho$. This signals  a change of orbital momentum
of the quark configurations participating in the one-gluon exchange
process. In order to obtain a nonzero result of the integration over
$\vec r$, either the initial or the final $\bar cc$ wave function
must contain a factor $\vec\nabla_r$, i.e. it must  be a $P$-wave.
Since we assume that the
Higgs is a scalar, its $\bar cc$ component must be in a $P$-wave
state, while the primordial $\bar cc$ in the projectile IC state
should be in an $S$-wave. This is vice versa for the proton $p_1$:
the final $|3q\ra$ system is  in an $S$-wave, but $\Psi_{3q}(\vec\rho)$ must
be a $P$-wave.

Notice that both the scalar Higgs and $\chi$ states may be produced from the same
IC component of the proton containing $S$-wave $\bar cc$.
However, the production of $J/\psi$, $\Upsilon$, $Z^0$ requires an IC
component containing a $P$-wave $\bar cc$, which is presumably
more suppressed.

The $P$-wave LC wave function of Higgs in impact parameter representation is given by
the Fourier transform of its Breit-Wigner propagator:
 \beq
H(\vec r) = i\,\frac{\sqrt{N_c\, G_F}}{2\pi}\, m_c\,
\bar\chi\,\vec\sigma\,\chi\, \frac{\vec
r}{r}\,
\left[\epsilon\,Y_1(\epsilon r)-{ir\over2}\,\Gamma_HM_H\,
Y_0(\epsilon r)\right]
\ .
 \label{1500}
 \eeq
Here $G_F$ is the Fermi constant, $\chi$ and $\bar\chi$ are the
spinors for $c$ and $\bar c$ respectively and
 \beq
\epsilon^2=\alpha(1-\alpha)M_H^2 - m_c^2\ ,
 \label{1505}
 \eeq
where $\alpha$ is the fraction of the LC momentum of the Higgs carried by the
$c$-quark. The functions $Y_{0,1}(x)$ in Eq.~(\ref{1500}) are the second order Bessel
functions and $\Gamma_H$ is the total width of the Higgs. Assuming $\Gamma_G\ll
M_H$, we neglect the second term in Eq.~(\ref{1500}).

The LC wave function Eq.~(\ref{1505}) assumes that the Higgs mass
is much larger than the quark masses, which is probably true for
charm and bottom. However, it is quite probable that for
top-antitop in the Higgs $2m_t>M_H$, then the wave function is
different,
 \beq
H_{\bar tt}(\vec r) = \frac{\sqrt{N_c\, G_F}}{2\pi}\, m_t\,
\bar\chi\,\vec\sigma\,\chi\, \frac{\vec
r}{r}\,\epsilon_t\,K_1(\epsilon_t r)\ ,
 \label{1506}
 \eeq
 where $K_1(x)$ is the modified Bessel function and
 \beq
\epsilon_t^2=m_t^2-\alpha(1-\alpha)M_H^2~~.
\label{1506a}
 \eeq

The probabilities computed from the wave functions Eqs.~(\ref{1500}) and (\ref{1506a})
require regularization in the ultraviolet limit
\cite{krt,e-loss}, as is the case of the $\bar qq$ wave function
of a transverse photon. Such wave functions are not solutions of the
Schr\"odinger equation, but are distribution functions for perturbative
fluctuations. They are overwhelmed by very heavy fluctuations with large
intrinsic transverse momenta, or vanishing transverse separations. Such
point-like fluctuations lead to a divergent normalization, but they do not
interact with external color fields, i.e.they are not observable. All the
expressions for any measurable quantity, including the cross section, are
finite.

As we have discussed, the IC wave function can be modeled as a nonperturbative $5-quark$
stationary state $|3qc\bar c\ra$, or as a perturbative
fluctuation $|3q\ra \to |3qc\bar c\ra$. Correspondingly, the $\bar
cc$ wave function within the Fock state will be assumed to be a linear combination of
nonperturbative and perturbative distribution amplitudes,
 \beq
\Psi_{\bar cc}(\vec r)=
\beta\,\Psi^{npt}_{\bar cc}(\vec r)+
\sqrt{1-\beta^2}\,\Psi^{pt}_{\bar cc}(\vec r)\ .
\label{1507}
 \eeq
 The parameter $\beta$ which controls the relation between the
nonperturbative and perturbative IC contributions, is such that $0\leq\beta\leq 1$.
The nonperturbative wave function should be an S-wave
solution of the Schr\"odinger equation. Assuming oscillator
potential we get,
 \beq
\Psi^{npt}_{\bar cc}(\vec r) = \sqrt{\frac{m_c\omega}
{2\pi}}\,\exp(-r^2\,m_c\,\omega/4)\ ,
 \label{1510}
 \eeq
 where $\omega$ is the oscillation frequency, as mentioned earlier.

Since the Higgs is produced from an $S$-wave $\bar cc$, the
perturbative distribution amplitude is ultraviolet stable and
can be normalized to one, in order to correspond to $P_c$ as a
probability to have such a charm quark pair in the proton
 \beq
\Psi^{pt}_{\bar cc}(\vec r) = \frac{m_c}
{\sqrt{\pi}}\,K_0(m_c r)\ .
 \label{1520}
 \eeq
Here the modified Bessel function $K_0(m_c r)$ is the Fourier transform of
the energy denominator associated with the $\bar cc$ fluctuation. We assume
the $c$ and $\bar c$ quarks to carry equal fractional momenta. For fixed
$\alpha_s$ the energy denominator governs the probability of the fluctuation
in momentum space, since perturbatively one treats the charm quarks as free
particles.

Now we can calculate the part of the matrix element in Eq.~(\ref{1400}) related to
Higgs production from the IC. We assume the initial $\bar cc$ wave function to be
a linear combination (\ref{1507}) of the nonperturbative, Eq.~(\ref{1510}) and
perturbative, Eq.~(\ref{1520}) wave functions, and the final state wave function
of the $\bar cc$ pair in the Higgs in the form (\ref{1500}). The result
reads,
 \beqn
&& \int d^2r\, H^\dagger(\vec r)\, e^{i\vec k\cdot\vec r/2}
\left(1-e^{-i\vec k\cdot\vec r}\right) \Psi_{\bar cc}(\vec r)=
\frac{4}{\sqrt{\pi}}\,
\frac{m_c^2}{M_H^2}\sqrt{N_cG_F}\
 \bar\chi\,\vec\sigma\,\chi\,\vec k
\nonumber\\ &\times&
\left[\beta\,\sqrt{\frac{\omega}{2m_c}} + \sqrt{1-\beta^2}\,
\ln\left(\frac{M_H}{2m_c}\right)\right]
\ .
\label{1550}
 \eeqn
 Here we have made use of
$M_H\gg m_c$ and expanded the exponentials $\exp(\pm
i\vec k\cdot\vec r)$ up to the first nonvanishing term.
We also dropped the integration over $\alpha$, assuming that the
$c$ and $\bar c$ in the IC component carry the same momentum, i.e.
$\Psi_{\bar cc}(r,\alpha) = \Psi_{\bar
cc}(r)\,\delta(\alpha-1/2)$. Correspondingly, we have fixed
$\epsilon=M_H/2$ and we will assume $\beta\ll 1$.

The result of integration in Eq.~(\ref{1550}) shows that the perturbative
contribution is quite enhanced relative to the nonperturbative term. First of all,
the enhancement by factor $\sqrt{2m_c/\omega}$ is due to the fact that the
projection to the point-like Higgs wave function is proportional to $\Psi_{\bar
cc}^{IC}(0)$, and the perturbative fluctuation has a smaller radius. Another
enhancement factor, $\ln(M_H/2m_c)$, is due to the long power tail in the momentum
distribution in the perturbative IC wave function, which the nonperturbative one has
a Gaussian cut off. Thus, the perturbative term in the matrix element
Eq.~(\ref{1550}) is relatively enhanced by one order of magnitude.

Notice that the nonrelativistic nonperturbative solution should not be used for
convolution with the highly perturbative Higgs wave function. The large transverse
momentum tail of the IC should be represented by the perturbative term in
Eq.~(\ref{1507}). Unfortunately, the normalization of such a perturbative tail is
unknown, and we normalize it to the IC weight $1\%$.

Enhancement of the perturbative intrinsic heavy flavor in Higgs production is
especially large for the top component. Using the IT wave function
Eq.~(\ref{1506}) we get
 \beqn
\int d^2r\, H_{\bar tt}^\dagger(\vec r)\, e^{i\vec k\cdot\vec r/2}
\left(1-e^{-i\vec k\cdot\vec r}\right)
\Psi^{pt}_{\bar tt}(\vec r) &=&
\frac{1}{\sqrt{\pi}}\,
\frac{m_t^2}{M_H^2}\sqrt{N_cG_F}\
 \bar\chi\,\vec\sigma\,\chi\,\vec k
\nonumber\\ &\times&
\left[1+\frac{1-\delta}{\delta}\,\ln(1-\delta)\right]\ ,
\label{1565}
 \eeqn
 where $\delta=M_H^2/4m_t^2$.

The proton is produced in a similar way from the P-wave $3q$ in
the projectile IC component. The exponentials, however, should not
be expanded, since the radius is not small. Therefore, using the
relation,
 \beq
\int\limits_0^{2\pi}d\phi\, \vec\rho\, \left(
e^{i\vec\rho\cdot\vec k/2} - e^{-i\vec\rho\cdot\vec k/2} \right) =
4\pi i\,\rho\, \frac{\vec k}{k}\, J_1(k\rho/2)\ , \label{1560}
 \eeq
 we get for the integral over $d^2\rho$ in Eq.~(\ref{1400}),
 \beqn
&&\int d^2\rho\, \frac{1}{\sqrt{\pi} R_{3q}}\,
\frac{\vec\rho}{\rho}\, e^{-\rho^2/2R_{3q}^2}
\left(e^{i\vec\rho\cdot\vec k/2} -
e^{-i\vec\rho\cdot\vec k/2}\right)
\frac{1}{\sqrt{\pi} R_p}\, e^{-\rho^2/2R_p^2}
\nonumber\\ &=&
i\,\frac{\sqrt{\pi}}{4}\,
\frac{{\cal R}^3\,\vec k}{R_p\,R_{3q}}\,
e^{-y}\left[J_0(y)-J_1(y)\right]\ ,
 \label{1570}
 \eeqn
 where ${\cal R}^2=2\, R_p^2\,R_{3q}^2/ (R_p^2+R_{3q}^2)$, and
$y={\cal R}^2\,k^2/32$.  For further estimates we assume that
$R_p=R_{3q}$, so ${\cal R}=R_p$. Since we assumed a meson-type
quark-diquark structure for the proton, the mean separation
$R_p^2=2\la r_{ch}^2\ra_p/3$. The transition proton form factor,
$\exp(-k^2{\cal R}^2/32)$, cuts off the integration over $d^2k$.

Now we are in a position to perform the last integration over
$\vec k$ in Eq.~(\ref{1400}),
 \beqn
&& \frac{d\sigma^{IC}(pp\to ppH)}{dx_2} =
\frac{32}{\pi^2}\,\frac{G_F\,P_{IC}(z)}{1- x_2}\,
\frac{m_c^4}{M_H^4}\, \frac{[\sigma^{\pi p}_{tot}(s^\prime)]^2}
{B(s^\prime)\,\la r^2_{ch}\ra_{p}}\,
\frac{\gamma^2(s^\prime)} {[2+2\gamma(s^\prime)+\gamma^2(s^\prime)]^3}
 \nonumber\\ &\times&
\left[1 +\frac{\la r^2_{ch}\ra_{p}} {16\,\la
r^2_{ch}\ra_{\pi}}\, {1\over \gamma(s^\prime)}\right]^2\,
\left[\beta\,\sqrt{\frac{\omega}{2m_c}} + \sqrt{1-\beta^2}\,
\ln\left(\frac{M_H}{2m_c}\right)\right]^2\ ,
\label{1600}
 \eeqn
 where $\gamma(s^\prime)=R_p^2/4R_0^2(s^\prime)$.

The cross section of Higgs production from the intrinsic bottom
has the same form as Eq.~(\ref{1600}), and we assume that the
weight of intrinsic heavy flavor scales as,
$P_{IQ}=P_{IC}m_c^2/m_Q^2$. However, as we found above, if the
Higgs mass is restricted by $M_H^2<4m_t^2$, production from
intrinsic perturbative top component of the proton has cross
section,
 \beqn
&& \frac{d\sigma^{IT}(pp\to ppH)}{dx_2} =
\frac{8}{\pi^2}\,\frac{G_F\,P_{IT}(z)}{1- x_2}\,
\frac{m_t^4}{M_H^4}\, \frac{[\sigma^{\pi p}_{tot}(s^\prime)]^2}
{B(s^\prime)\,\la r^2_{ch}\ra_{p}}\,
\frac{\gamma^2(s^\prime)} {[2+2\gamma(s^\prime)+\gamma^2(s^\prime)]^3}
 \nonumber\\ &\times&
\left[1 +\frac{\la r^2_{ch}\ra_{p}} {16\,\la
r^2_{ch}\ra_{\pi}}\, {1\over \gamma(s^\prime)}\right]^2\,
\left[1+\frac{1-\delta}{\delta}\,\ln(1-\delta)\right]^2\ .
\label{1750}
 \eeqn

\subsection{Feynman \boldmath$x^H_F$ distribution of Higgs particles}

The $x_F^H$ distribution of the cross section Eqs.~(\ref{1600})-(\ref{1750}) is
related to the LC wave function $\Psi_{IC}(R,z)$ of the system $3q-\bar cc$,
namely to the function $P_{IC}(z)$ defined in Eq.~(\ref{1460}). The momentum
fraction $z$ is related to $x_{1,2}$ and $x_F^H$ by Eq.~(\ref{1405}). The shape of
$P_{IC}(z)$ strongly correlates with the origin of IC, a nonperturbative component
of the proton wave function or a perturbative fluctuation.

\subsubsection{Nonperturbative IC}

In principle one can construct hadronic LC wave function by
diagonalizing the LC Hamiltonian. Here we will use the method of
Ref.~\cite{terentiev} for the Lorentz boost of the wave function,
which is supposed to be known in the hadron rest frame. The Lorentz
boost generates higher particle number quantum fluctuations which
are missed by this procedure; however this method works  well in
known cases~ \cite{hikt,kth}, and even provides a nice cancelation
of large terms violating the Landau-Yang theorem~ \cite{kt}.

We assume that the rest frame IC wave function has the oscillatory
form (in momentum space),
 \beq
\tilde\Psi_{IC}(\vec Q,z) = \sqrt{P_{IC}(z)}
\left(\frac{1}{\pi\omega\mu}\right)^{3/4} \exp\left(-\frac{\vec
Q^2}{2\omega\mu}\right)\ .
 \label{2000}
 \eeq
 Here $\omega$ stands for the oscillator frequency and
$\mu=M_{\bar cc}M_{3q}/(M_{\bar cc}+M_{3q})$ is the reduced mass
of the $\bar cc$ and $3q$ clusters. For further estimates we use
$M_{\bar cc}=3\GeV$ and $M_{3q}=1\GeV$, although the latter could
be heavier, since it is the $P$-wave.

To express the 3-vector $\vec Q$ by the effective mass of the
system, $M_{eff}=\sqrt{\vec Q^2 +M_{\bar cc}^2}+ \sqrt{\vec Q^2
+M_{3q}^2}$, one can switch to the LC variables, $\vec Q$ and
$z$,
 \beq
M_{eff}^2=\frac{Q^2}{z(1-z)}+\frac{M_{\bar cc}^2}{z}+
\frac{M_{3q}^2}{1-z}\ . \label{2020}
 \eeq
 Then the longitudinal component $Q_L$ in the exponent in (\ref{2000})
reads,
 \beq
Q_L^2=\frac{M_{eff}^2}{4}+ \frac{(M_{\bar
cc}^2-M_{3q}^2)^2}{4M_{eff}^2} - \frac{M_{\bar cc}^2+M_{3q}^2}{2}
- Q^2\ , \label{2040}
 \eeq
 and the LC wave function acquires the form,
 \beq
\Psi_{IC}(Q,z) = K\,\sqrt{P_{IC}(z)}\,
\exp\left\{-\frac{1}{8\omega\mu}\left[M_{eff}^2 + \frac{(M_{\bar
cc}^2-M_{3q}^2)^2}{M_{eff}^2} \right]\right\}\ , \label{2060} \eeq
 where
 \beqn
K^2 &=& \frac{1}{8Q_L}\,
\left(\frac{1}{\pi\omega\mu}\right)^{3/2}\,
\exp\left(\frac{M_{\bar cc}^2+M_{3q}^2}{2\omega\mu}\right)\,
\left[1-\frac{(M_{\bar cc}^2-M_{3q}^2} {M_{eff}^4}\right]
\nonumber\\ &\times& \left[\frac{Q^2(2z-1)}{z^2(1-z)^2} -
\frac{M_{\bar cc}^2}{z^2}+ \frac{M_{3q}^2}{(1-z)^2}\right]~.
\label{2080} \eeqn

\begin{figure}[htb]
\begin{center}
\leavevmode {\epsfxsize=8.0cm \epsffile{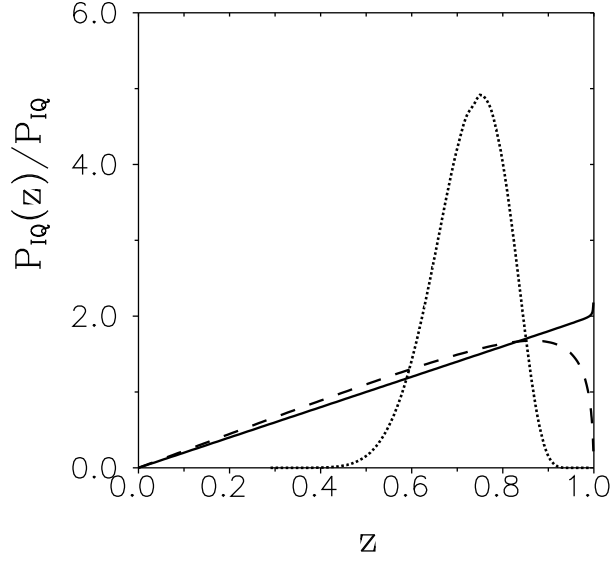}}
\end{center}
 \caption[*]{\baselineskip 1pt The distribution of produced Higgs
particles over the fraction of the proton beam momentum. The dotted,
dashed and solid curves correspond to Higgs production from
nonperturbative IC ($\beta=1$), perturbative IC ($\beta=0$) and IT,
respectively.}
 \label{fi:3}
\vspace*{-1.5ex}
 \end{figure}

Now we can calculate the $z$-dependence of the function
$P_{IC}(z)$ defined in Eq.~(\ref{1460}), which controls the $x_1$
dependence of the cross section,
 \beq
\frac{P_{IC}(z)}{P_{IC}}= \
\frac{1}{\sigma^{IC}(pp\to ppH)}\,
\frac{d\sigma^{IC}(pp\to ppH)}{dx_1}
= \frac{1}{P_{IC}} \int
d^2Q\,\left|\Psi_{IC}(Q,z)\right|^2~.
 \label{2100}
 \eeq
 This function is plotted in Fig.~\ref{fi:3}.
The distribution sharply picks at $z\approx 0.75$, as one could expect,
since the IC pair is heavy and should carry the main fraction of
the proton momentum. Note that at high energies, in particular at
LHC, the momentum fraction $z$ coincides with the Feynman $x_H$ of
the Higgs particle, with a high accuracy $\sim M_H^2/s$.

\subsubsection{Perturbative intrinsic heavy flavors}

The light-cone wave function of a perturbative fluctuation $p\to |3q\bar
QQ\ra$
in momentum representation is controlled by the energy denominator,
 \beq
\Psi_{IQ}(Q,z,\kappa)\propto
\frac{z(1-z)}{Q^2+z^2m_p^2+M^2_{\bar QQ}(1-z)}\ .
 \label{1800}
 \eeq
 Momentum $\vec Q$ was defined in Fig.~\ref{fi:2} and Eq.~(\ref{15}).  The effective
mass of the $\bar QQ$ depends on the intrinsic transverse momentum of the $\bar
QQ$
pair, $M_{\bar QQ}^2=4(\kappa^2+m_Q^2)$. It is controlled by the convolution of
the
IC $\bar QQ$ wave function with the $P$-wave $\bar QQ$ wave function in the Higgs
and the one-gluon exchange amplitude (see Fig.~\ref{fi:2}), which has the form,
 \beqn
&& \int\limits_{0}^{\infty} d\kappa^2\,
\Psi_{IQ}(Q,z,\kappa)\left[H_{\bar QQ}(\vec\kappa+\vec k/2)-
H_{\bar QQ}(\vec\kappa-\vec k/2)\right]
\nonumber \\ &\propto& z(1-z)\,
\frac{\ln\left[\frac{\left|M_H^2-4m_Q^2\right|(1-z)}
{Q^2+4m_Q^2(1-z)+m_p^2z^2}\right]}
{M_H^2(1-z)+Q^2+m_p^2z^2}\ .
 \label{1820}
 \eeqn
 This expression peaks at $1-z\sim m_p/M_H$, therefore the logarithmic factor
hardly varies as function of $Q^2$ which is restricted by the proton
form factor. Making use of this we perform integration in
Eq.~(\ref{2100}) and arrive at the following $z$-distribution,
 \beq
\frac{P_{IQ}(z)}{P_{IQ}}=
Nz(1-z)\,
\frac{\left\{\ln\left[\frac{\left|M_H^2-4m_Q^2\right|(1-z)}
{4m_Q^2(1-z)+m_p^2z^2}\right]\right\}^2}
{M_H^2(1-z)+m_p^2z^2}\ ,
 \label{1830}
 \eeq
 where $N$ is a constant normalizing to one the integral over $z$.  The
corresponding $z$-distributions for charm and top are shown in Fig.~\ref{fi:3} by
dotted (dashed) and solid curves respectively.

\subsection{Energy dependence}

One can integrate in Eqs.~(\ref{1600})-(\ref{1750}) over $x_2$ using relation
(\ref{1405}). Since the momentum distribution of Higgs produced from the
nonperturbative IC sharply peaks at $z=z_0=0.75$, one can replace
$P_{IC}(z)\Rightarrow \delta(z-z_0)\,P_{IC}$. With a reasonable accuracy we can
fix $z$ at the same value for the perturbative case and heavier flavors too, which
is justified by the rather mild dependence on $\tilde s$ of other factors in
Eqs.~(\ref{1600})-(\ref{1750}).

Since at high energies $z\approx x_H\approx 1-x_1$, performing
integration in Eq.~(\ref{1600}) one arrives at,
 \beqn
&& \sigma^{IC}(pp\to ppH) =
\frac{32}{\pi^2}\,\frac{G_F\,P_{IC}}{z_0}\,
\frac{m_c^4}{M_H^4}\, \frac{[\sigma^{\pi p}_{tot}(\tilde s)]^2}
{B(\tilde s)\,\la r^2_{ch}\ra_{p}}\,
\frac{\gamma^2(\tilde s)} {[2+2\gamma(\tilde s)+\gamma^2(\tilde
s)]^3} \nonumber\\ &\times&
\left[1 +\frac{\la r^2_{ch}\ra_{p}} {16\,\la
r^2_{ch}\ra_{\pi}}\, {1\over \gamma(\tilde s)}\right]^2\,
\left[\beta\,\sqrt{\frac{\omega}{2m_c}} + \sqrt{(1-\beta^2)}\,
\ln\left(\frac{M_H}{2m_c}\right)\right]^2\ ,
\label{1700}
 \eeqn
 where $\tilde s=s\,z_0\,M_0^2/M_H^2$.
Analogous expression should be valid for Higgs production from
intrinsic bottom. For top quark in the proton we use
Eq.~(\ref{1750}) which leads to,
 \beqn
&& \sigma^{IT}(pp\to ppH) =
\frac{8}{\pi^2}\,\frac{G_F\,P_{IT}}{z_0}\, \frac{m_t^4}{M_H^4}\,
\frac{[\sigma^{\pi p}_{tot}(\tilde s)]^2} {B(\tilde s)\,\la
r^2_{ch}\ra_{p}}\, \frac{\gamma^2(\tilde s)} {[2+2\gamma(\tilde
s)+\gamma^2(\tilde s)]^3} \nonumber\\ &\times& \left[1 +\frac{\la
r^2_{ch}\ra_{p}} {16\,\la r^2_{ch}\ra_{\pi}}\, {1\over
\gamma(\tilde s)}\right]^2\,
\left[1+\frac{1-\delta}{\delta}\,\ln(1-\delta)\right]^2\ .
\label{1705}
 \eeqn

Notice that function $\gamma(\tilde s)$ increases with energy as
$s^{0.28}$, and such a steep rise of the denominator in Eq.~(\ref{1700})
is not compensated by the rise of the total cross section in the
numerator. Therefore, the diffractive cross sections,
Eqs.~(\ref{1700})-(\ref{1705}),
turns out to decrease at asymptotic energies approximately as inverse
energy. This unexpected result may be interpreted as follows. The
source of the falling energy dependence is the steep rise with energy
of the mean transverse momentum of gluons as is given by the
unintegrated gluon density Eq.~(\ref{1410}), $\la k^2\ra
=4/R_0^2(x)\propto (s/M_H^2)^{0.28}$. Also the integral over $k^2$ of
the distribution (\ref{1410}) rises with energy, and its value at $k=0$
is steeply falling. The rise comes for large transverse momenta which,
however, are cut off by the nucleon form factor Eq.~(\ref{1510}). This
is why the diffractive cross section (\ref{1700}) is steeply falling.
Indeed, without this form factor, for instance in the reaction $pp\to
HXp$, the cross section would rise as $(s/M_H)^{0.7}$.

Nevertheless, at the energy of LHC, $\sqrt{s}=14\TeV$, the effective
energy is
rather low, $\sqrt{\epsilon} s=120\GeV$ (we assume
$M_H=100\GeV$) the cross section still rises with energy. Indeed,
$R_0^2=0.36\fm^2$, so $\gamma(\epsilon s)=0.55$ is still rather small
at this energy, and the cross sections Eqs.~(\ref{1700})-(\ref{1705}) rise as,
 \beq
 \sigma^{IQ}(pp\to ppH)_{LHC} \propto
\left(\frac{s}{M_H^2}\right)^{0.6}\ .
 \label{1620}
 \eeq
 However, at much higher energies the energy dependence will switch to
a steeply falling one. Besides, absorptive or unitarity corrections are
known to slow down the rise of the cross sections.

\subsection{Absorptive corrections}\label{abs}

The amplitude of any off-diagonal large rapidity gap process is
subject to unitarity or absorptive corrections, which have the
intuitive meaning of a survival probability of the participating
hadrons. To include these corrections one should replace the
diffractive amplitude as,
 \beq
f^{pp}_{sd}(b,s)\Rightarrow f^{pp}_{sd}(b,s)\,\left[1-\Im
f^{pp}_{el}(b,s)\right]\ , \label{306}
 \eeq

The data for elastic $pp$ scattering show that the partial amplitude
$f^{pp}_{el}(b,s)$ is constant energy at small impact
parameters $b\to 0$, while rising as function of energy at large
$b$ \cite{amaldi,kp1,k3p}. This is usually interpreted as a
manifestation of saturation of the unitarity limit, $\Im
f^{pp}_{el}\leq 1$. Indeed, this condition imposes a tight
restriction at small $b$, where $\Im f^{pp}_{el}\approx 1$,
leaving almost no room for further rise.
We will treat the Pomeron as a Regge pole without
unitarity corrections:
 \beq
\Im f^{pp}_{el}(b,s)= \frac{\sigma^{pp}_{tot}(s)} {4\pi
B^{pp}_{el}(s)}\ \exp\left[-\frac{b^2}{2 B^{pp}_{el}(s)}\right]\ ,
\label{307}
 \eeq
 where $\sigma^{pp}_{tot}(s)=21.8\mb\times(s/M_0^2)^{\epsilon}$, and $\epsilon=0.08$; $B^{pp}_{el}(s)=B^0_{el}+2\,\alpha^\prime_{\Pom}\,
\ln(s/M_0^2)$ with $B^{0}_{el}=7.5\GeV^{-2}$. Due to the accidental closeness of $2\alpha_{\Pom}^\prime/B^0_{el}= 0.067$ and $\epsilon$,
the pre-exponential factor in (\ref{307}) hardly changes with energy even
without unitarity corrections.  It is demonstrated in Ref.~\cite{k3p} that not only at $b=0$,
but in the whole range of impact parameters, the model
Eq.~(\ref{307}) describes correctly the energy dependence of the partial amplitude $f^{pp}_{el}(b,s)$.

Thus we arrive at the absorption corrected cross section,
 \beqn
&& \tilde\sigma^{IQ}(pp\to ppH) = \sigma^{IQ}(pp\to ppH)\,
\nonumber\\ &\times&
\left\{1-{1\over{\pi}}\,\frac{\sigma^{pp}_{tot}(s^\prime)}
{B(s^\prime)+2B^{pp}_{el}(s^\prime)} + \frac{1}{(4\pi)^2}\,
\frac{\left[\sigma^{pp}_{tot}(s^\prime)\right]^2}
{B^{pp}_{el}(s^\prime) \left[B(s^\prime)+B^{pp}_{el}(s^\prime)
\right]}\right\}\ .
 \label{308}
 \eeqn
 This is not a severe suppression even at the energy of LHC, where the
absorptive factor is $0.2$.

\begin{figure}[htb]
\begin{center}
\hspace*{1.5cm}
\leavevmode {\epsfxsize=8.0cm
\epsffile{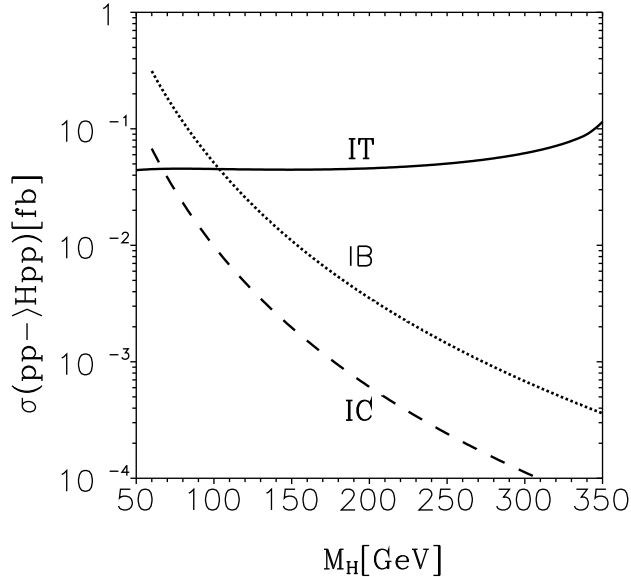}}
\vspace*{-0cm}
\hspace*{4cm}
\end{center}
 \caption[*]{\baselineskip 1pt
 The cross section of the reaction $pp\to Hp+p$ as function of the Higgs mass.
 Contributions of IC (dashed) of IB (dotted) and IT (solid). }
 \label{fi:4}
 \vspace*{-1.5ex}
 \end{figure}

Including the absorptive corrections we calculated the total cross sections for
diffractive Higgs production, $pp\to Hpp$, from the IQ components. The results at the
energy of LHC, $\sqrt{s}=14\TeV$, are plotted as function of Higgs mass in the
Fig.~\ref{fi:4}. We assume a perturbative origin for all intrinsic components, a
$1/m_Q^2$ scaling for their weights, and $1\%$ probability of IC for $\beta=0$ in
Eq.~(\ref{1700}).
 Note that the contributions of the intrinsic charm and bottom falls steeply with the
mass of the Higgs in accordance with Eq.~(\ref{1700}). The contribution of the intrinsic top rises with $M_H$ unless $M_H > 2m_t \approx 350\GeV$, then the cross section starts
falling.


\section{Further possibilities to get a larger cross section}

\subsection{Direct production of Higgs from a colorless IQ}

\begin{figure}[tbh]
\includegraphics{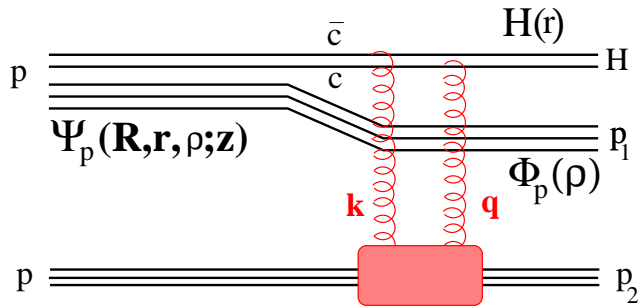}
\begin{center}
\vspace{6.5cm}
\parbox{13cm}
{\caption[Delta]
 {Higgs production via Pomeron exchange }
 \label{dif-direct}}
\end{center}
 \end{figure}

A heavy flavor $\bar QQ$ pair in the IQ component of the proton may be found
in a colorless state. In this case the Higgs particle can be
produced directly from this pair via Pomeron exchange as is shown
in Fig.~\ref{dif-direct}.
 We consider the example of intrinsic charm of a nonperturbative origin throughout
this section.

At first glance one may think that this channel is less suppressed
by powers of the Higgs mass compared to the mechanism presented in
Fig.~\ref{fi:2}. Indeed, the integration over $\vec k$ does not
have the upper cutoff imposed by the proton form factor in the
previous case;  therefore it may compensate two powers of $M_H$ in
the amplitude. This analysis is correct. Nevertheless the amplitude turns
out to be more suppressed than in the diagram Fig.~\ref{fi:2}.

The diffractive amplitude $A(x_2,\vec p_1,\vec p_2)$ is
proportional to the matrix element of the dipole cross section
$\sigma_{\bar qq}(r)$ between the initial $\bar cc$ wave function
$\Psi_{\bar cc}(r)$ and the distribution amplitude of the $\bar
cc$ in the Higgs,
 \beq
A(x_2,\vec p_1,\vec p_2) \propto \int d^2r\, H^\dagger(\vec r)\,
\sigma_{\bar qq}(r)\, \Psi_{\bar cc}(r)\ .
 \label{2200}
 \eeq
 This factor contains all the dependence on the Higgs mass. To estimate
it one can use a Gaussian shape for both $\Psi_{\bar cc}(r)$ and
$H(\vec r)$. Then one finds $A\propto \sqrt{m_c\omega}/M_H^3$.
A more refined calculation confirms this,
 \beqn
\int d^2r\, H^\dagger(\vec r)\, \sigma_{\bar qq}(r)\, \Psi_{\bar
cc}(r) &=& \frac{3\sigma_0(s)\sqrt{2\pi N_cG_F}}
{2m_c\omega\,R_0^2(s)}\ {\rm U}\left({3\over2},0,\frac{M_H^2}
{4m_c\omega}\right) \nonumber\\ &=&
\frac{12\sigma_0(s)}{R_0^2(s)}\, \sqrt{2\pi N_cG_F}\,
 \frac{\sqrt{m_c\omega}}{M_H^3}
\ , \label{2300}
 \eeqn
 where the initial state $\bar cc$ pair is assumed to be in a $P$-wave.
Here $U(a,b,x)$ is the confluent hypergeometric function, and we
use its asymptotic behavior at $x\gg1$,
 \beq
U(a,b,x)=x^{-a}+O(x^{-a-1})\ . \label{2310}
 \eeq

Notice that a convolution similar to Eq.~(\ref{2200}) also defines
the scale dependence of the amplitude of photoproduction of heavy
quarkonia, which also behaves like $(Q^2+M^2)^{-3}$.
Thus, the complementary mechanism of diffractive Higgs production,
besides the smaller probability to find the colorless IC
component, is additionally suppressed by $1/M_H^2$ compared to the
dominant mechanism depicted in Fig.~\ref{fi:2}. Therefore, this
contribution can be safely neglected.

\subsection{Nuclear enhancement}

The produced Higgs is supposed to escape detection and to be identified
only by using the missing mass spectrum.  One may also consider the same
reaction on a bound proton in $pA$ collisions  where the nuclear debris
spectators flying in the same direction as the Higgs. The nuclear
enhancement in this case is not as large as one could naively expect.
The reason is that absorptive corrections are stronger than
those considered above in Sect.~\ref{abs} for the case of $pp$ collisions.
The survival probability represented by the last factor in Eq.~(\ref{308}) can be evaluated within the Glauber
approximation  for $pA$ collisions as
 \beq
Z_{eff} = \frac{Z}{A}\int
d^2b\,T_A(b)\,e^{-\sigma^{pp}_{in}T_A(b)}\ ,
 \label{2330}
 \eeq
 where $T_A(b)=\int_{-\infty}^{\infty}dz\,\rho_A(b,z)$ is the nuclear
thickness function at impact parameter $\vec b$, and $\rho_A(b)$ is the
nuclear density.  We assume here that diffractive recoil neutrons cannot
be detected; otherwise the factor (\ref{2330}) should be multiplied by
$A/Z$.

The nuclear enhancement for lead according to (\ref{2330}) is rather
mild, of about an order of magnitude, since $Z_{eff}\approx 2.5$ should be
compared with the suppression factor $0.2$, for $pp$ collisions
calculated in Sect.~\ref{abs}. Gribov corrections \cite{gribov} are known
to make nuclei more transparent, therefore they may substantially
increase the survival probability factor \cite{mine,kps}. If we employ
the simplest quadratic dependence of the dipole cross section
$\sigma_{\bar qq}(r) \propto r^2,$ then the nuclear enhancement is
considerably larger, a factor of about $50$.

\section{Conclusions}

The key assumption underlying our analysis of high $x_F$ Higgs hadroproduction is the presence of
intrinsic heavy flavor   $|uud Q \bar Q>$ Fock
components in the proton bound-state wave function.
Such quantum
fluctuations are in fact rigorous consequences of QCD.  The
the probability  for intrinsic heavy quark on the heavy quark mass falls as $\Lambda^2_{QCD}\over M_Q^2$ in non-abelian theories and can be
computed from the operator product expansion ~\cite{Franz:2000ee}.
In such Fock states the
heavy quarks $Q$ and $\bar Q$  carry the highest light-cone momentum fractions.
Thus although they have small probability, intrinsic heavy quark Fock states are highly efficient in transferring the momentum
of the proton to the momenta of particles in the final-state, especially  heavy quarkonium and the Higgs which can sum the momenta of both the $Q$ and $\bar Q$.   It is thus interesting and important that measurements of the production of heavy quarkonium at high $x_F$ as well as other heavy hadrons such as the $\Lambda_b$ and $\Lambda_c$  be carried out at RHIC, the Tevatron, as well as the LHC in order to test this novel feature of QCD.

 As we have reviewed in section 3, there is substantial but not conclusive phenomenological evidence for intrinsic charm at the $1\%$ probability level in the proton.  It is thus particularly important to have  measurements of  the charm and bottom structure functions in deep inelastic lepton-proton scattering over the full range of $x_{bJ}$. One must allow for intrinsic sea components at any scale $Q_0$ when parameterizing the proton's structure functions, since the intrinsic Fock states are responsible for the $\bar u(x) \ne \bar d(x)$, $s(x) \ne \bar s(x) $ asymmetries, as well as the high-$x$ $c(x)$ and $b(x)$ distributions.
There are also important nuclear and heavy quark threshold effects related to intrinsic charm which can be tested at lower energy fixed-target facilities such as JLAB, GSI-FAIR, and J-PARC~\cite{Brodsky:2004ah}.

As we have emphasized here, the materialization of intrinsic heavy
flavor states in the proton leads to Higgs production in the
proton fragmentation region: this includes inclusive production $
pp \to  H  X,$ singly diffractive production $p p \to p + H X$ and
exclusive diffractive production $ p p \to p + H +p$, reactions
which should be considered in addition to the conventional central
rapidity production processes. The fractional momentum
distribution for a Higgs produced by combining the momenta of both
heavy quarks in the  IQ Fock states is presented in
Fig.~\ref{fi:3}.  As seen in the figure, the Higgs can be produced
with momentum fractions as large as  $x_F \sim 0.8$ or even
higher.  One also produce the Higgs inclusively from leading-twist
PQCD processes such as $ g c \to H c$ and $g b \to H b$ where the
high momentum of one intrinsic heavy quark is transferred to the
Higgs.

We have focused in this paper on diffractive exclusive Higgs
production $p p \to p + H +p ,$ since in principle, only the
final-state protons need to be measured and the Higgs can be
reconstructed from the missing mass distribution.   We note,
however, that detecting the diffractive signal $p p \to p + H +p$
poses new challenges: When the Higgs is produced at large $x_F$,
one of the final-state protons will be need to be detected at a
small momentum fraction $\sim 1-x_F$, which is outside of the
usual acceptance of forward proton detectors.

The underlying color structure of the intrinsic Fock state and the
gauge theory properties of the two-gluon exchange mechanism for
high energy diffraction play key roles in the physics of the
exclusive diffractive Higgs hadroproduction process. The main
result of our analysis, the cross sections given in
Eqs.~(\ref{1700},\ref{1705}), demonstrates that the heavier the
intrinsic heavy quark, the larger is the cross section for the
doubly diffractive reaction. It rises with $m_Q$ linearly if the
heavy quarks are confined by a potential, and is presented in
Fig.~\ref{fi:4} if the $\bar QQ$ appear in the proton as a
perturbative fluctuation. The production cross section also
steeply rises with energy $\propto s^{0.7}$, which is
characteristic for the energy dependence of hard reactions.
Absorptive correction slow down this rise and eventually stop it
at very high energies, above the energy range of LHC.
Asymptotically, in the Froissart regime, this cross section is
expected to fall.  Numerical predictions for diffractive Higgs
production from IC, IB and IT components are shown in
Fig.~\ref{fi:4}. The cross section will be further enhanced from
possible Fock states of the proton containing supersymmetric
partners of quarks or gluons.  We also have discussed  a potential
increase in the rate for such reactions using proton-nucleus
collisions.

\bigskip

{\bf Acknowledgments:}
Work supported in part by the Department of Energy
under contract number DE-AC02-76SF00515, by Fondecyt (Chile) grant
numbers 1030355 and 1050519, by DFG (Germany) grant PI182/3-1 and
by the cooperation program Ecos-Conicyt C04E04 between France and
Chile.

We are thankful to Fred Goldhaber, Paul Hoyer  and Alexander Tarasov for informative and helpful discussions.

\end{document}